\documentclass[aps,prl,twocolumn,superscriptaddress,showpacs]{revtex4}
\usepackage{graphicx}
\usepackage{amsmath}
\usepackage{amssymb}
\input{epsf}
\usepackage{epsfig}

\newcommand{\rem}[1]{}

\begin{document}

\title{Entanglement Echoes in Quantum Computation}
\author{Davide Rossini}
\affiliation{Center for Nonlinear and Complex Systems, Universit\`a degli 
Studi dell'Insubria, Via Valleggio 11, 22100 Como, Italy}
\author{Giuliano Benenti}
\email{giuliano.benenti@uninsubria.it}
\homepage{http://www.unico.it/~dysco}
\affiliation{Center for Nonlinear and Complex Systems, Universit\`a degli 
Studi dell'Insubria, Via Valleggio 11, 22100 Como, Italy}
\affiliation{Istituto Nazionale per la Fisica della Materia, 
Unit\`a di Como, Via Valleggio 11, 22100 Como, Italy}
\author{Giulio Casati}
\email{giulio.casati@uninsubria.it}
\affiliation{Center for Nonlinear and Complex Systems, Universit\`a degli 
Studi dell'Insubria, Via Valleggio 11, 22100 Como, Italy}
\affiliation{Istituto Nazionale per la Fisica della Materia, 
Unit\`a di Como, Via Valleggio 11, 22100 Como, Italy}
\affiliation{Istituto Nazionale di Fisica Nucleare,
Sezione di Milano, Via Celoria 16, 20133 Milano, Italy}
\date{September 19, 2003}

\begin{abstract} 
We study the stability of entanglement in a quantum computer
implementing an efficient quantum algorithm, which simulates 
a quantum chaotic dynamics. 
For this purpose, we perform a forward-backward evolution of 
an initial state in which two qubits are in a maximally 
entangled Bell state. If the dynamics is reversed after an
evolution time $t_r$, there is an echo of the entanglement 
between these two qubits at time $t_e=2t_r$. 
Perturbations attenuate the pairwise entanglement echo and generate 
entanglement between these two qubits and the other qubits of the 
quantum computer. 
\end{abstract}
\maketitle

The development of new techniques which could enhance the
reliability of quantum computation is intrinsically connected 
with the study of its stability.
Every physical implementation of a quantum computer will have to 
deal with errors, due to the coupling with the environment or to an 
imperfect control of the computer hardware.
Therefore an accurate study of the stability of a quantum computer, 
while it is running quantum algorithms, is demanded \cite{pazzurek}.

Entanglement is arguably the most peculiar feature of 
quantum systems, with no analog in classical mechanics.
Furthermore, it is an important physical resource, which is 
at the basis of many quantum information protocols, including
quantum cryptography \cite{cryptography} and teleportation 
\cite{teleportation}. 
For any quantum algorithm operating on pure states, the presence of 
multipartite (many-qubit) entanglement is necessary to achieve an 
exponential speedup over classical computation \cite{jozsa}. 
Therefore the ability to control entangled states is one of the 
basic requirements for constructing quantum computers.

In this paper, we introduce a suitable method to characterize the 
stability of pairwise entanglement in quantum computation, by
considering the echo of an initially maximally entangled pair of qubits.
Namely, we assume that our quantum computer is initially in the state
$\vert\psi_0\rangle=\vert\Phi_B\rangle\otimes\vert\chi\rangle$.
Here, the first two qubits are prepared in a maximally 
entangled Bell state ($\vert\Phi_B\rangle$), while the other 
$n_{q}-2$ qubits are set in a pure state ($\vert\chi\rangle$), 
and they are completely disentangled from the Bell pair 
($n_q$ denotes the total number of qubits in the quantum computer).
The state $|\psi_0\rangle$ first evolves according to the given 
quantum algorithm, described by the unitary evolution operator 
$\hat{\cal U}$. 
Then we invert the sequence of quantum gates that implement 
this algorithm, that is we apply $\hat{\cal U}^\dagger$. In the 
ideal case we would reconstruct the initial state, since 
$\hat{\cal U}^\dagger \hat{\cal U}\vert\psi_0\rangle=|\psi_0\rangle$. 
However, due to noise and imperfections, the initial state 
$\vert\psi_0\rangle$ is not exactly recovered. 
In particular, the first two qubits are no longer in a Bell state, 
and therefore their pairwise {\it entanglement echo} is reduced.
Conversely, this pair of qubits becomes entangled with the other 
qubits, thus generating multipartite entanglement.
In this paper, we study numerically the attenuation 
of the pairwise entanglement echo in a quantum computer implementing 
an efficient quantum algorithm which simulates quantum chaotic
dynamics.
We point out that the entanglement echo simulations discussed 
in the following are close in spirit to the spin echo experiments 
in many-body quantum systems in the presence of perturbations 
\cite{spinecho}.

We study the entanglement echo for the quantum algorithm 
simulating the sawtooth map dynamics \cite{bcms01}.  
The sawtooth map is a periodically driven dynamical system, 
described by the Hamiltonian
\begin{equation}
H(\theta,n,\tau) = \frac{n^{2}}{2} - \frac{k (\theta - \pi)^{2}}{2} 
\sum_{j=-\infty}^{+\infty} \delta(\tau-jT),
\end{equation}
where $(n,\theta)$ are conjugated action-angle variables 
($0 \leq \theta < 2\pi$).
The time evolution $\tau \to \tau+T$ of this system is classically 
described by the map
$\bar{n} = n + k (\theta - \pi)$, and $\bar{\theta} = 
\theta + T \bar{n}$,
where the bars denote the variables after one map iteration.
By rescaling $n \to p=Tn$, one can see that classical dynamics 
depends only on the parameter $K=kT$. The classical motion is stable 
for $-4 \leq K \leq 0$ and completely chaotic for $K<-4$ and $K>0$.
The quantum evolution in one map iteration is described by the 
unitary operator $\hat{U}$:
\begin{equation}
\vert \bar{\psi} \rangle = \hat{U} \vert \psi \rangle =
e^{-iT\hat{n}^{2}/2} \, e^{ik(\hat{\theta} -\pi)^{2}/2} 
\vert \psi \rangle,
\label{quantmap} 
\end{equation}
where $\hat{n} = -i \partial/\partial \theta$  
(we set $\hbar=1$).
The classical limit is obtained by taking $k \to \infty$ and 
$T \to 0$, keeping $K=kT$ constant.
We study map (\ref{quantmap}) on the torus $0 \leq \theta <2\pi$, 
$-\pi \leq p < \pi$. With a $n_{q}$ qubits quantum computer, 
we can simulate the quantum dynamics of the sawtooth map with 
$N=2^{n_q}$ levels, and we set $T=2\pi/N$.
The effective Planck's constant of the quantum system is 
$\hbar_{\mathrm{eff}} \sim 1/N$ and 
the classical limit corresponds to $n_q\to \infty$ 
($\hbar_{\mathrm{eff}} \to 0$) \cite{bcms01}.
We focus on the case $K=5$, which corresponds to the chaotic 
regime.

It is convenient to simulate map (\ref{quantmap}) by means of 
the forward-backward Fourier transform between $\theta$ and $n$ 
representations.
While the classical fast Fourier transform requires 
$O(N\log_{2}N)$ operations, an efficient quantum algorithm has been
found \cite{bcms01}, which uses the quantum Fourier 
transform and simulates (\ref{quantmap}) in 
$O(n_{q}^{2}=(\log_2 N)^2)$ elementary
quantum gates per map iteration.
Moreover, all $n_q$ qubits are used in an optimal way, that is 
no extra work space qubits are required. 
In this way interesting physical phenomena, like dynamical localization
\cite{bcms03}, cantori localization, and anomalous diffusion 
could be simulated already with less than $10$ qubits. Therefore the
quantum sawtooth map represents an interesting testing ground 
for quantum computation, and it is important to understand the 
limits to the quantum computation of this model due to noise
and imperfections. 

To compute the entanglement echo, we start from the initial state
\begin{equation}
\vert\psi_0\rangle=
\vert\Phi_B\rangle\otimes \vert\chi\rangle =
\frac{1}{\sqrt{2}} 
\left( \vert 00 \rangle + \vert 11 \rangle \right) 
\otimes \vert 00 \ldots 0 \rangle
\label{psiin} 
\end{equation}
and we perform a forward evolution of the 
quantum sawtooth map (\ref{quantmap}) up to time $t=t_r$, that is 
$\hat {\cal U} = \hat{U}^{t_r}$ (the discrete time $t=\tau/T$ denotes 
the number of map iterations). 
Then we compute the time reversal evolution up to the 
{\it echo time} $t_e=2 t_r$ (namely, $\hat{\cal U}^\dagger=
(\hat{U}^\dagger)^{t_r}$).
Our algorithm can be decomposed into single-qubit Hadamard gates 
and two-qubit controlled-phase shift gates \cite{bcms01}. 
In particular, the Hadamard gate can be written 
as $\hat{\bf n}_{\scriptscriptstyle H} \cdot \boldsymbol{{\sigma}}$, 
where $\hat{\bf n}_{\scriptscriptstyle H}=(1/\sqrt{2},0,1/\sqrt{2})$, 
and $\boldsymbol{\sigma}=(\sigma_x,\sigma_y,\sigma_z)$,
the $\sigma_i$'s being the Pauli matrices.
Due to the imperfect control of the quantum system during the quantum 
computation, the initial state is not perfectly recovered.
In this paper, we only deal with unitary errors, modeled by noisy gates. 
We assume that errors tilt the rotation axis  
$\hat{\bf n}_{\scriptscriptstyle H}$ 
by an angle randomly fluctuating in the interval $[-\epsilon,\epsilon]$.
In the noisy controlled-phase shift gates, random phases of 
amplitude inside the interval $[-\epsilon,\epsilon]$ are added. 
We assume that the errors affecting two consecutive quantum gates 
are completely uncorrelated.

Since we consider unitary errors, the state $\vert\psi(t)\rangle$ 
of the quantum computer at any time $t$ is still a pure state. 
By tracing $\vert \psi(t) \rangle$ over all the qubits,
except those initially prepared in a Bell state, we obtain the
reduced density matrix 
$\rho_{12} (t)=
\mathrm{Tr}_{3, \ldots, n_{q}} \, 
\left( \vert {\psi(t)} \rangle \langle {\psi(t)} \vert \right) $.
Note that, in general, qubits $1$ and $2$ are no longer disentangled 
from the other qubits of the quantum computer, and therefore 
$\rho_{12}$ is a mixed state.
We evaluate the entanglement of formation $E(t)$ of the state 
$\rho_{12}$ following Ref.~\cite{wootters}. 
First of all we compute the concurrence, defined as 
$C=\max ( \lambda_{1} - \lambda_{2} - 
\lambda_{3} - \lambda_{4} , 0 )$,
where the $\lambda_{i}$'s are the square roots of the eigenvalues 
of the matrix $R= \rho_{12} \tilde{\rho}_{12}$, in decreasing order.
Here $\tilde{\rho}_{12}$ is the spin flipped matrix of $\rho_{12}$, 
and it is defined by $\tilde{\rho}_{12}= 
(\sigma_{y} \otimes \sigma_{y}) \, \rho_{12}^{\star} \, 
(\sigma_{y} \otimes \sigma_{y})$ 
(note that the complex conjugate is taken in the computational
basis $\{ \vert 00 \rangle, \vert 01 \rangle, \vert 10 \rangle, 
\vert 11 \rangle \}$).
Once the concurrence has been computed, entanglement is obtained
as $E= h((1+\sqrt{1-C^{2}})/2)$, where $h$ is the binary
entropy function: $h(x)=-x\log_{2}x-(1-x)\log_{2}(1-x)$.
We also compute the Von Neumann entropy 
$S(t)=-\mathrm{Tr} \, [ \rho_{12}(t) \log_{2} \rho_{12}(t)]$
of the reduced density matrix $\rho_{12}$.
This quantity measures the entanglement between the qubits $1$ and $2$ 
and the other $n_q-2$ qubits of the quantum computer. 
In particular, we compute the entanglement echo $E(t_e)$ and 
the Von Neumann entropy $S(t_e)$ at the echo time $t_e$.

\begin{figure}
\centerline{\epsfxsize=8cm \epsffile{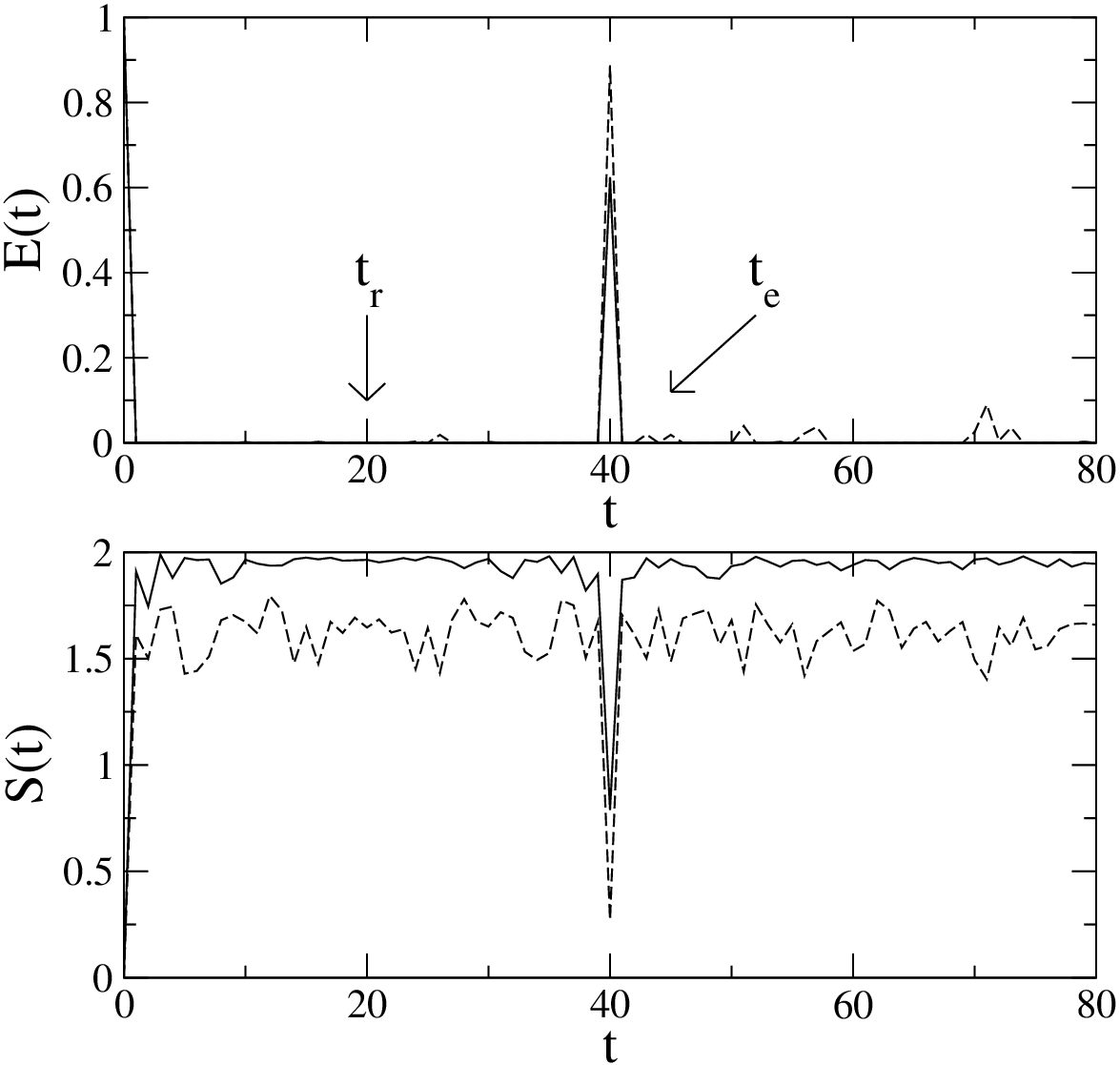}}
\caption{Entanglement echo in a noisy 
quantum computer implementing the sawtooth map algorithm
in the chaotic regime at $K=5$, with $n_q=5$ qubits
(dashed line) and $n_q=8$ qubits (solid line), and 
perturbation strength $\epsilon=10^{-2}$.
We start from the initial state (\ref{psiin}), and, 
from $t=0$ to $t_r=20$ a forward evolution of the sawtooth map 
is applied. After that time, we invert the dynamics. The echo 
occurs at time $t_e=2t_r=40$.
Top: entanglement of qubits $1$ and $2$.
Bottom: entanglement between these two qubits and the other ones.}
\label{fig1}
\end{figure}

A typical numerical simulation of the entanglement echo 
is shown in Fig.~\ref{fig1}.
The upper part shows the behavior of the pairwise entanglement
of the two qubits initially prepared in a Bell state;
in the lower part we plot the Von Neumann entropy of this
two-qubit subsystem.
The dynamics completely destroys the initial pairwise entanglement,
which is partially recovered only at the echo time $t_e$
\cite{notechaos}. 
Instead, the Von Neumann entropy quickly reaches the  
saturation value.
These results can be understood as follows: 
in few map iterations ($t\sim 1$), the chaotic dynamics
transforms the initial state $\vert\psi_0\rangle$ into an 
ergodic state.
We can expand the state $\vert\psi(t)\rangle$ at time $t$ 
over the computational basis: 
$\vert \psi(t) \rangle =  
\sum_{\alpha_{1} \ldots  \alpha_{n_{q}}}
c_{\alpha_{1}, \ldots , \alpha_{n_{q}}} (t)\vert \alpha_{1}
\ldots  \alpha_{n_{q}} \rangle$,
where $\alpha_i=0,1$ for $i=1,...,n_q$. 
Since $\vert\psi(t)\rangle$ is ergodic,
the coefficients $c_{\alpha_{1}, \ldots , \alpha_{n_{q}}} (t)$
have random phases and amplitudes $\vert c_{\alpha_{1}, \ldots ,
\alpha_{n_{q}}} \vert \sim 1/\sqrt{N}$ 
(to assure wave function normalization).
For an ergodic system, the reduced density matrix of a 
two-qubit subsystem is essentially diagonal.
Indeed, the diagonal matrix elements are given by
$(\rho_{12})_{\alpha_1,\alpha_2;\alpha_1,\alpha_2} = 
\sum_{\alpha_{3},...,\alpha_{n_q}} \vert
c_{\alpha_{1}, \ldots,  \alpha_{n_{q}}}\vert^2$,
and their value is $\approx 1/4$, since they are given 
by the sum of $N/4$ positive terms, whose value is $\sim 1/N$. 
The off-diagonal matrix elements of $\rho_{12}(t)$ are instead given 
by the sum of $N/4$ terms of amplitude $1/N$ and random phases. 
Hence their value is $O(1/\sqrt{N})$. 
For such a nearly diagonal density matrix $\rho_{12}(t)$,  
the entanglement of formation can be analytically computed and 
we find that $E=0$.
This means that chaotic dynamics quickly destroys the entanglement 
of any two-qubit subsystem, as shown in Fig.~\ref{fig1}. 
Under the hypothesis that the wave function is ergodic, 
it is also possible to compute analytically the Von Neumann 
entropy $S$ of the qubits $1$ and $2$.
After averaging over noise realizations, one has 
$S\approx 2-8/(N\ln 2)$ \cite{page,lakshminarayan}, 
in good agreement with our numerical data.
This value is close to the maximum possible entropy of the 
two-qubit subsystem, $S_{\rm max}=2$.
It is interesting to note that it is possible to invert the 
quantum dynamics after long times (much longer than the times 
for relaxation to statistical equilibrium) and recover the 
initial out of equilibrium state. 
This is a clear demonstration of the stability of the quantum motion 
in contrast to the high instability of the classical chaotic 
motion \cite{casati}.

\begin{figure}
\centerline{\epsfxsize=8.cm \epsffile{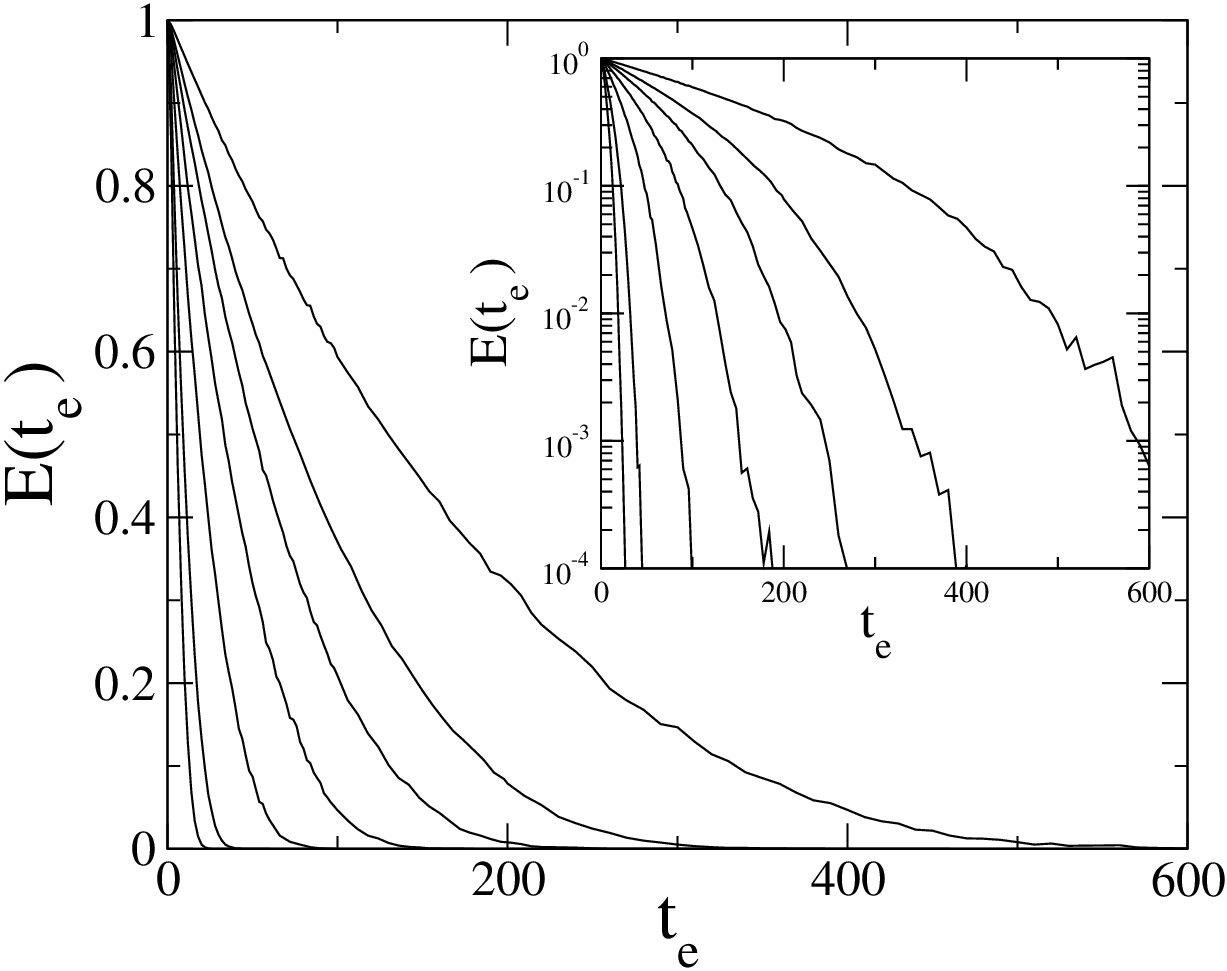}}
\caption{Attenuation of the entanglement echo of a Bell pair 
in the sawtooth map, at $K=5$, $n_q=7$, and, from right 
to left, $\epsilon= 7.5 \times 10^{-3}, 10^{-2}, 1.2\times 10^{-2},
1.5 \times 10^{-2}, 2 \times 10^{-2}, 3 \times 10^{-2}, 
4 \times 10^{-2}$.
Here and in the following figures data are averaged over 400 
runs with different noise realizations.
Inset: semilogarithmic plot of the same curves.}
\label{fig2}
\end{figure}

\begin{figure}
\centerline{\epsfxsize=8.cm \epsffile{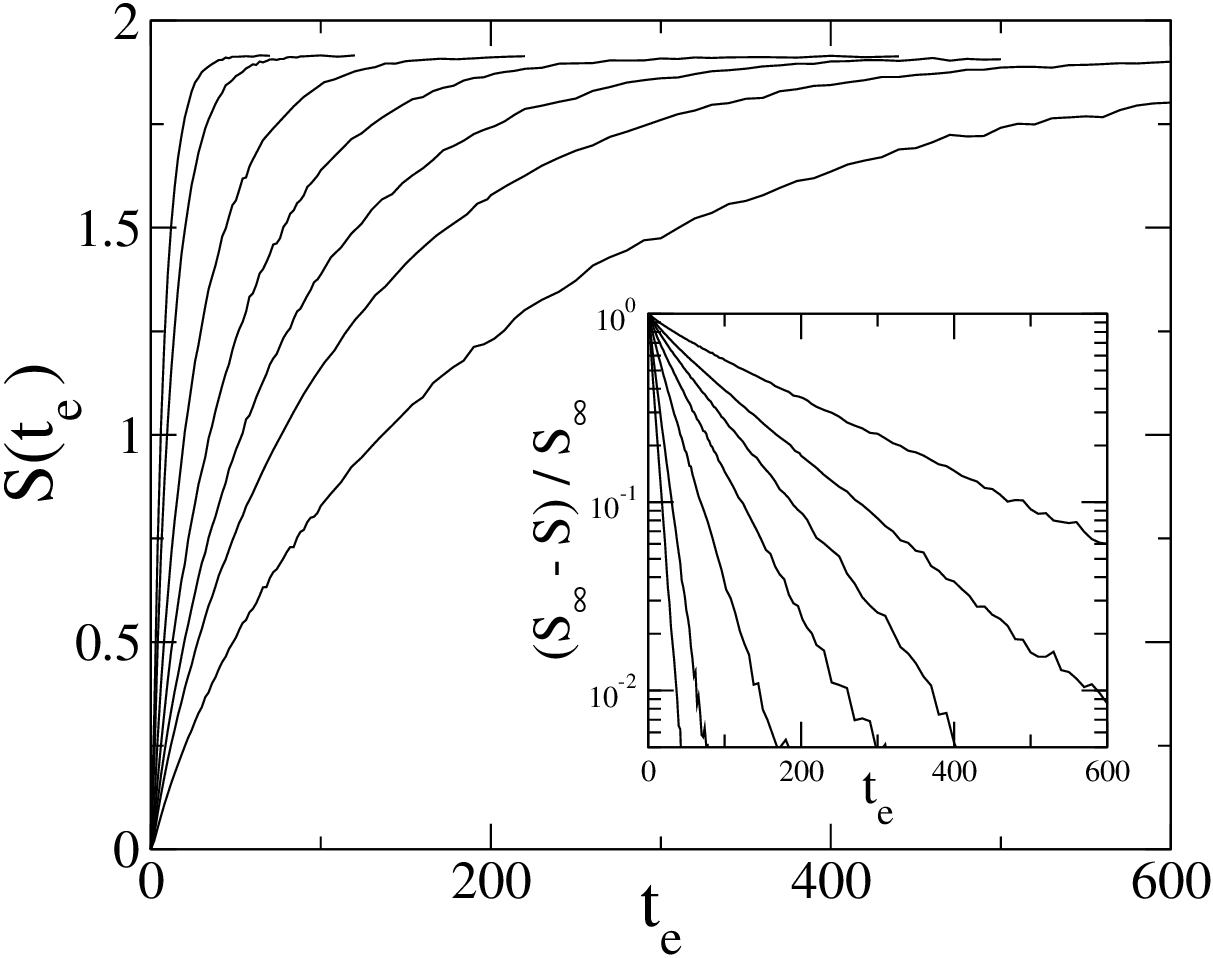}}
\caption{Von Neumann entropy of the two 
initially entangled qubits as a function of the echo time $t_e$,
with same parameter values as in the previous figure.
Inset: approach to the saturation value $S_{\infty}$ for the 
same curves.}  
\label{fig3}
\end{figure}

We now focus on the stability of the entanglement echo at time $t_e$.
The numerically computed entanglement echo and Von Neumann entropy 
at the echo time $t_e$ are shown, for different noise strengths, 
in Figs.~\ref{fig2} and \ref{fig3}, respectively.
Fig.~\ref{fig2} shows that noisy gates attenuate the entanglement 
echo, and, if the time $t_e$ is long enough, completely destroy it. 
This can be explained by noticing that unitary errors 
transform the echo state into a state which becomes 
closer to an ergodic state, as $t_e$ increases. As we have 
seen above, an ergodic state of the whole quantum computer 
implies a pairwise entanglement $E=0$ \cite{lewenstein}. 

\begin{figure}
\centerline{\epsfxsize=8.cm \epsffile{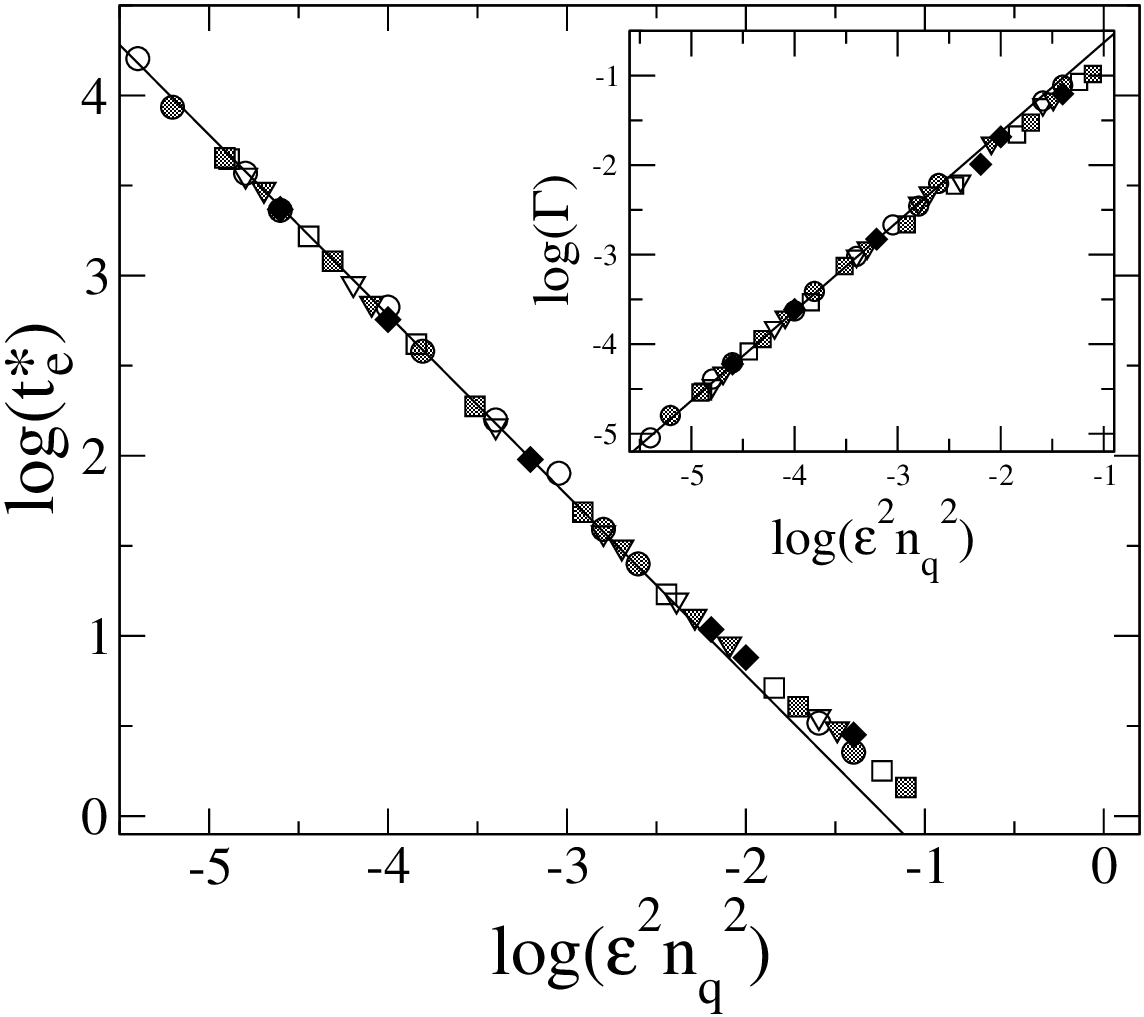}}
\caption{Time scale $t_e^\star$ for the decay of the entanglement echo in 
the sawtooth map at $K=5$, for different strengths and number of qubits:
$n_q=4$ (empty circles), $5$ (filled circles), $6$ (empty squares),
$7$ (filled squares), $8$ (empty triangles), $9$ (filled triangles), 
and $10$ (diamonds). Straight line: $t_{e}^\star=A /n_{q}^{2} 
\epsilon^{2}$, with the fitting constant $A\approx 6.04\times 10^{-2}$.
Inset: rate $\Gamma$ of the approach to equilibrium of the Von Neumann
two-qubit reduced entropy. The straight line gives 
$\Gamma=B \epsilon^2 n_{q}^2$, with $B\approx 2.34$.
Logarithms are decimal.} 
\label{fig4}
\end{figure}

The decay of the entanglement echo can be understood 
by considering that each noisy gate 
transfers a probability of order $\epsilon^{2}$ from the 
ideal state to all other states.  
Since there are no correlations between consecutive noisy gates,
the population of the initial state decays exponentially, and 
we can write the echo state as follows:
\begin{equation} 
\vert \psi(t_e) \rangle \approx  
e^{-C\epsilon^{2} n_{g} t_e/2} |\psi_0\rangle +
\sum_{\boldsymbol{\alpha}\ne
\boldsymbol{\alpha}_{\scriptscriptstyle A},
\boldsymbol{\alpha}_{\scriptscriptstyle B}} 
a_{\boldsymbol{\alpha}}(t_e) |\boldsymbol{\alpha}\rangle, 
\label{psie}
\end{equation}
where $C$ is a constant to be determined numerically,
$n_g t_e$ is the total number of gates required to perform 
the echo experiment ($n_g=3n_q^2+n_q$ being the number of gates 
per map iteration), and the sum runs over all
the states $\vert{\boldsymbol{\alpha}}\rangle=\vert\alpha_1...
\alpha_{n_q}\rangle$ of the computational basis, except for the 
two states involved in the initial wave vector 
$\vert\psi_0\rangle$ 
($\vert{\boldsymbol{\alpha}_{\scriptscriptstyle A}}\rangle=
\vert 000...0\rangle$ and 
$\vert{\boldsymbol{\alpha}_{\scriptscriptstyle B}}\rangle=
\vert 110...0\rangle$).
Given the complexity of the dynamics simulated by the
quantum algorithm for the sawtooth map, it is reasonable to assume 
that the coefficients $a_{\boldsymbol{\alpha}}$ have random signs 
and amplitudes of the order of 
$\sqrt{(1-e^{-C\epsilon^2 n_g t_e})/(N-2)}$ (to assure that 
$\vert\psi(t_e)\rangle$ has unit norm).  
We can compute the entanglement echo $E(t_e)$ from 
the expression (\ref{psie}) for the echo wave function.
For $\epsilon^{2} n_{g} t_e \ll 1$, it turns out that 
$E(t_e) \approx 1 - (3/2\ln 2) C \epsilon^{2} n_{g} t_e$.
Therefore the entanglement echo is stable up to time 
$t_e\propto 1/(\epsilon^2 n_g)\propto 1/(\epsilon^2 n_q^2)$.
This theoretical estimate is confirmed by Fig. \ref{fig4}, in which 
we plot the characteristic time scale $t_e^\star$ for the decay of 
the entanglement echo, defined by the condition $E(t_e^\star)=c$ 
(we take $c=0.9$).
It is clearly seen that $t_e^\star \propto n_{q}^{-2} \epsilon^{-2}$.
Therefore noisy gates degrade the entanglement echo after
a number $n_e^\star$ of elementary gates 
which is {\it independent of the number of qubits}  
($n_e^\star=n_g t_e^\star\propto \epsilon^2$). 

The Von Neumann entropy $S(t_e)$ of the reduced two-qubit 
subsystem at the echo time is shown in Fig.~\ref{fig3}.
It saturates, for sufficiently long echo times, to the value 
$S_\infty \approx 2 - 8/(N \ln 2)$, as expected for an 
ergodic state of the quantum computer. 
We can compute the approach to equilibrium from 
Eq.~(\ref{psie}), that gives $S_\infty-S(t_e)\propto
\exp(-2C\epsilon^2 n_g t)$. This theoretical prediction 
is borne out by our numerical data shown in 
Figs.~\ref{fig3}-\ref{fig4}.
The inset of Fig.~\ref{fig3} shows that the approach to  
the asymptotic value is exponential. Indeed, we have 
$S(t_e) \approx S_\infty (1-e^{-\Gamma t_e})$.  
In Fig.~\ref{fig4} (inset) we plot the rate $\Gamma$
for different number of qubits and perturbation strengths. 
We see that $\Gamma \propto \epsilon^{2} n_{q}^2$.

It is interesting to compare the entanglement echo decay with 
the decay of the fidelity, which is the usual tool used to 
characterize the stability of quantum computation 
\cite{pazzurek,bcms01}.
The fidelity $f$ at time $t_e$ is defined as 
$f(t_e)=\vert\langle \psi(t_e) | \psi_0\rangle\vert^2$, and
Eq.~(\ref{psie}) implies that $f(t_e)\approx
\exp(-C \epsilon^2 n_g t_e)$.
This is in agreement with our numerical data (not shown here). 
Therefore the decay of the fidelity, the entanglement echo and 
the approach to equilibrium for the reduced Von Neumann entropy
take place {\it in the same time scale} 
$\propto 1/(\epsilon^2 n_g)$ \cite{footnote}.

In summary, we have proposed a suitable method, the entanglement
echo, to study the stability of entanglement under perturbations.
We have shown that noise destroys the entanglement of a pair
of qubits and produces entanglement between these two qubits and 
the other qubits of the quantum computer. 
We point out that, since the entanglement can be measured 
experimentally in an efficient way \cite{horodecki}, entanglement
echo experiments analogous to the numerical simulations discussed 
in this paper could be implemented in quantum processors with a small 
number of qubits (4-10) and a few hundreds of gates. These experiments 
are close to present capabilities \cite{NMR1,NMR2,ions} and would bring 
new insights in our understanding of the limits to quantum computation 
due to decoherence and imperfections.

\begin{acknowledgments}

This work was supported in part by the EC contracts 
IST-FET EDIQIP and RTN QTRANS, the NSA and ARDA under
ARO contract No. DAAD19-02-1-0086, and the PRIN 2002 
``Fault tolerance, control and stability in
quantum information processing''.

\end{acknowledgments}

\bibliographystyle{prsty}

\begin{thebibliography}{99}
\bibitem{footnote} The value of the constant $C$ that appears in
Eq.~(\ref{psie}) can be determined from our numerical data for
the fidelity decay, the entanglement echo decay (Fig.~\ref{fig4}),
and the relaxation to equilibrium (inset of Fig.~\ref{fig4}). 
We get $C\approx 0.28$, $0.26$, and $0.39$, respectively.
The rather good agreement between these values is a further
confirmation of the validity of our theoretical analysis.
\end{thebibliography}

\end{document}